\documentclass[prd,superscriptaddress,nofootinbib,showkeys,showpacs,11pt]{revtex4-1}
\newcommand{\im}{{\mbox{Im}\,}}

\usepackage{amsmath}
\usepackage{amsfonts}
\usepackage{amssymb}
\usepackage{graphicx}
\usepackage{epstopdf}
\usepackage{dcolumn}
\usepackage{bm}
\usepackage{color}
\usepackage{dsfont}

\begin{document}

\def\eq{\begin{eqnarray}}
\def\en{\end{eqnarray}}

\title{Bound states on the lattice with partially twisted boundary conditions}

\author{D. Agadjanov}

\affiliation{Helmholtz-Institut f\"ur Strahlen- und Kernphysik (Theorie) and
Bethe Center for Theoretical Physics,
Universit\"at Bonn,
D-53115 Bonn, Germany
}

\affiliation{St. Andrew the First-Called Georgian University of the Patriarchate of Georgia, Chavchavadze Ave. 53a, 0162, Tbilisi, Georgia}

\author{F.-K. Guo}
\author{G. R\'{\i}os}
\author{A. Rusetsky}

\affiliation{Helmholtz-Institut f\"ur Strahlen- und Kernphysik (Theorie) and
Bethe Center for Theoretical Physics,
Universit\"at Bonn,
D-53115 Bonn, Germany
}

\begin{abstract}

We propose a method to study the nature of exotic hadrons by determining
the wave function renormalization constant $Z$ from lattice simulations.
It is shown that, instead of studying the volume-dependence of the spectrum,
one may investigate the dependence of the spectrum on the twisting angle,
imposing twisted boundary conditions on the fermion fields on the lattice.
In certain cases, e.g., the case of the $DK$ bound state which is addressed
in detail, it is demonstrated that the partial twisting is equivalent to
the full twisting up to exponentially small corrections.

\end{abstract}



\maketitle

\section{Introduction}

The search for the exotic states
(tetraquarks, hybrids, hadronic molecules, etc)
in the observed hadron spectrum has been a subject of
both theoretical and experimental investigations for decades.
The exact pattern, how these states emerge, should be strictly
determined by the underlying theory and
should therefore contain important information about the behavior of QCD
at low energies. In practice, however, extracting such information from
the data  encounters certain
challenges, which are in part of a conceptual nature. In the present
paper we wish to focus exactly on this issue.

In general, a state is called ``exotic'' if its quark content does not
correspond to the ``standard'' constellation given by the non-relativistic
quark model ($q\bar q$ for mesons and $qqq$ for baryons). Consequently,
one needs to use a particular model as a reference point
to define how the exotic states are meant (note that the very notion of
constituent quarks is, strictly speaking, model-dependent).
Putting it differently, one has to {\it agree} on certain {\it criteria} 
formulated in terms of certain hadronic {\it observables}:
if these observables are measured, or calculated on the lattice, and the results
do not follow the pattern predicted by the quark model, this then should be
interpreted as a signature for exotica.

A standard example for the exotic state candidates is given by the scalar
nonet with the masses around 1~GeV. As it is well known, the observed
mass hierarchy
in this nonet is reversed as compared to, e.g., the pseudoscalar or
vector multiplets. Such a mass ordering is counter-intuitive from the point
of view of the naive quark model, but can be easily understood, if the scalar
mesons were interpreted as tetraquark states (see,
e.g.,~\cite{Jaffe:1976ig,Black:1998wt,Achasov:2002ir,Pelaez:2004xp}).
This is, however, not the only possible interpretation.
In Refs.~\cite{Weinstein:1982gc,Oller:1997ng,Oller:1998hw}, the $a_0(980)$ and
$f_0(980)$ were considered as hadronic molecules, whereas
in Refs.~\cite{Oller:1998zr}
these states were described as a combination of a bare pole and
the rescattering contribution.
In the J\"ulich meson-exchange model, the $f_0(980)$ appears to be a
 bound $K\bar K$ state, whereas the $a_0(980)$ is a dynamically
generated threshold effect \cite{Janssen:1994wn}. Similar conclusions were inferred
in Ref.~\cite{Guo:2011pa} from the calculations in
the unitarized ChPT with explicit resonance states.
Finally, the investigations carried out within the
framework of QCD sum rules are also
 indicative of the non-$q\bar q$ nature of $a_0(980)$~\cite{sumrules}.
Given these multiple interpretations, it is natural to look for the clear-cut
criteria based on the {\it observables} in order to minimize the
{\it model-dependence} of the statements about the nature of the hadronic states
in question.

In fact, such criteria are known for quite some time already.  The
``pole counting'' method, considered in
Refs.~\cite{Morgan:1992ge,Tornqvist:1994ji}, relates the number of the $S$-matrix
poles near threshold to the molecular nature of the states corresponding to these
poles. Namely, it has been argued that the loosely bound states of hadrons
(hadronic molecules) correspond to a single pole, whereas the poles corresponding
to the tightly bound quark states (of standard or exotic nature) always come in
pairs. A closely related criterion goes under the name of Weinberg's
compositeness condition~\cite{Weinberg}, which uses the quantity called the
wave function
renormalization constant $Z$, where $0\leq Z\leq 1$, to differentiate between the loosely
bound states and tight QCD composites, the values $Z\simeq 0$ corresponding
to the molecular states and {\it vice versa.} The application of these
methods for the analysis of the data on scalar mesons are considered in
Refs.~\cite{Morgan:1993td,Baru:2003qq,Baru:2004xg,Hanhart:2006nr,Hanhart:2007cm,Hyodo:2011qc}, and the recent review on the subject may be found in Ref.~\cite{Hyodo:2013nka}.
Moreover, {\it theoretically,} one may study the dependence of the pole positions
on the number of the colors $N_c$ (see Refs.~\cite{Nebreda:2011cp,Guo:2011pa,Pelaez:2006nj}) or the quark masses (Refs.~\cite{Hanhart:2008mx,Pelaez:2010fj,Bernard:2010fp,Albaladejo:2012te}). From the above studies, one can judge about the precise
structure of these states beyond the simple alternative between a molecule and
a tight quark composite.

Recent years have seen a renewed interest in the field, which is partly related
to the progress in the lattice calculations of the QCD spectrum
at the quark masses close to the physical values. It should be realized
that the lattice studies have powerful tools at their disposal to analyze the
nature of the states that emerge in QCD. Apart from the information about the
dependence of the spectrum on quark masses, a valuable information
comes from the volume dependence of the calculated spectrum as well as its
dependence on the twisting angle in case of twisted boundary
conditions, see Refs.~\cite{Suganuma:2005ds,MartinezTorres:2011pr,Sekihara:2012xp,Ozaki:2012ce,Albaladejo:2013aka}. Note that all this information is obtained from the
first-principle calculations on the lattice and is thus in principle 
devoid of any
model-dependent input.

In this paper we investigate the nature of the scalar states in the sector with
one charm quark that is a natural generalization of our treatment of the light
scalar mesons. We mainly focus on the case of the $D_{s0}^*(2317)$
meson~\cite{experiment},
albeit the formalism, which we develop here, can be straightforwardly applied
to the other cases where a bound state close to the elastic threshold
emerges (note that, in this paper, we do not consider the generalization
of the approach
to the inelastic case. This forms a subject of a separate investigation.).
The $D_{s0}^*(2317)$ does not fit very nicely to the quark-model picture, and
its structure is still debated, see, e.g., Ref.~\cite{Zhu:2007wz} for a recent
review. The molecular picture, due to the closeness of the $DK$ threshold
and a large coupling to the $DK$ channel looks most promising among
other alternatives. It would be highly desirable to verify this conjecture in
a model-independent manner, on the basis of the lattice calculations.
To this end, one may use the fact that the dependence of the bound-state
energy on the kaon mass is very different for a molecule and a standard
quark-model state, see Ref.~\cite{Cleven}. Another possible method to address
this issue has
been described, e.g., in Refs.~\cite{MartinezTorres:2011pr,Sekihara:2012xp},
where the authors propose to study the volume-dependence of the spectrum
in order to apply the Weinberg's compositeness criterion on the lattice.

The exploratory study of light pseudoscalar mesons $(\pi,K)$ of $(D,D_s)$ in
full lattice QCD has been carried out in Refs.~\cite{Liu:2012zya,Moir:2013yfa,prelovsek}.
In some isospin channels the study is plagued by the presence of disconnected
contributions. The implementation of the method from
  Refs.~\cite{MartinezTorres:2011pr,Sekihara:2012xp}, which implies carrying
out calculations at different volumes, could be therefore quite expensive.
In this paper we propose an alternative, which requires calculations at
one volume, albeit with twisted boundary conditions.
Moreover, we show that, in the study of $D_{s0}^*(2317)$, one may use
{\it partially twisted} boundary conditions, despite the fact that the
quark annihilation diagrams are present. The method used in the proof is
the same as in Ref.~\cite{Agadjanov:2013kja}. Generally,
one may expect that the
simulations with partially twisted boundary conditions could be
less expensive than working at different volumes, while they provide us
the same information about the nature of the bound states in question.

This article is organized as follows. In Sect.~\ref{compositeness}, we
briefly review Weinberg's argument for the compositeness of particles.
In Sect.~\ref{sec:mass-shift} we describe the procedure of
extraction of the parameter $Z$ from the data with twisted boundary
condition.
Further, in Sect.~\ref{models} we use some models and produce synthetic lattice
data in order to check the procedure of the extraction in practice.
The error analysis has also been carried out.
Separately,
in Sect.~\ref{partially}, we discuss the use of the partially twisted boundary
conditions and show that they are equivalent to the full twisting in our case.
Sect.~\ref{conclusions} contains our conclusions.

\section{Compositeness of bound states}
\label{compositeness}

As mentioned before, in view of the plethora of candidates of exotic hadrons,
it is very important to make model-independent statements on the nature of
these states. Model-independence requires that we can only
study the physical observables which can be defined in terms of the matrix
elements between asymptotic states. In particular, we would like to ask a
question, whether a given particle, corresponding to the $S$-matrix pole,
can be regarded as ``elementary'' or rather as a bound state
(molecule) of other hadrons.
The central place in this identification belongs to
 the so-called wave function renormalization constant $Z$, which
has been used to distinguish composite particles from elementary
ones since the early
1960's~\cite{Howard,Vaughan,Salam:1962ap,Weinberg:1962hj,Weinberg,WeinbergBook}.
To see its role, we will first discuss a non-relativistic quantum mechanical
system, following the discussion of Ref.~\cite{Weinberg}.

In this section, we will restrict our discussion to the infinite volume. Let us
consider a two-body system with a Hamiltonian
$    \mathcal{H} = \mathcal{H}_0 + V$,
where $\mathcal{H}_0$ is the free Hamiltonian, and $V$ specifies the
interaction. Both $\mathcal{H}$ and $\mathcal{H}_0$ have a continuum spectrum.
Let us assume that there is a bound state solution of the 
Schr\"{o}dinger
equation
with a binding energy $E_B$,
\begin{equation}
    \mathcal{H} |B\rangle = -E_B |B\rangle,
    \label{eq:sch}
\end{equation}
and $\mathcal{H}_0$ also has a discrete spectrum which are the bare elementary
particles. For simplicity, we will assume that there is only one such state,
denoted by $|B_0\rangle$. In the Hilbert space spanned by the eigenstates of
the free Hamiltonian, the completeness relation is thus given by
\begin{equation}
    1 = |B_0\rangle\langle B_0| + \int\! \frac{d^3 \vec{q} }{(2\pi)^3} \,
|\vec{q}\,\rangle \langle \vec{q}\,| \quad\text{with}\quad
\mathcal{H}_0|\vec{q}\,\rangle = \frac{\vec{q}^{\,2} }{2\mu} |\vec{q}\,\rangle,
\end{equation}
where $\mu=m_1 m_2/(m_1+m_2)$ is the reduced mass. Thus, the probability for the
physical state $|B\rangle$ overlapping with the
elementary state $|B_0\rangle$ which, by definition, equals to $Z$,
is given by
\begin{eqnarray}
    Z=\big| \langle B_0|B\rangle \big|^2 = 1 - \int\! \frac{d^3 \vec{q}
}{(2\pi)^3} \, \big| \langle \vec{q}\,|B\rangle \big|^2 = 1 - \int\!
\frac{d^3 \vec{q} }{(2\pi)^3} \, \frac{\big| \langle \vec{q}\,|V|B\rangle
\big|^2 }{ \left[E_B + \vec{q}^{\,2}/(2\mu)\right]^2 },
\label{eq:BB0}
\end{eqnarray}
where Eq.~\eqref{eq:sch} is used. The quantity $ 1 - \big| \langle
B_0|B\rangle \big|^2$ then describes the probability of the physical state not
being the elementary state or finding the physical state in the two-particle
state. In other words, $Z\simeq 1$ corresponds to a mostly
elementary state whereas a state with $Z\simeq 0$ can be interpreted as
a predominately molecular one.

In general, the above integral depends on the matrix element
$\langle \vec{q}\,|V|B\rangle$, which is not directly measurable. However,
for loosely  bound states, the quantity $Z$ can be
related to the
observables. Consider, for instance, an
S-wave bound state with a small binding energy. The binding
energy should be much smaller
than the inverse of the range of forces so that the matrix
element $\langle \vec{q}\,|V|B\rangle$ can be approximated by a constant
$g_\text{NR}$. We get from Eq.~\eqref{eq:BB0}
\begin{equation}
    g_\text{NR}^2 = (1-Z) \frac{2\pi}{\mu^2}\sqrt{2\mu E_B}.
    \label{eq:gnr}
\end{equation}
Note that, in the past, this equation
has been often applied to distinguish composite particles from elementary
ones, see e.g.~\cite{Weinberg,Baru:2003qq,Guo:2013zbw,Hyodo:2013nka}.
The non-relativistic coupling constant $g_\text{NR}^2$ coincides with the
residue of the
non-relativistic scattering
matrix at the bound state pole. This can be immediately
seen, considering the Low equation 
\begin{equation}
    t(E) = \frac{g_\text{NR}^2 }{ E+E_B+i\epsilon } 
+ \int\!\frac{d^3\vec q}{ (2\pi)^3 } \frac{ | t(E_q)|^2}{E-E_q+i\epsilon }
\label{eq:low}
\end{equation}
in the vicinity of the pole~\cite{Weinberg,WeinbergBook}.
Here, $E_q = \vec{q}^{\,2}/(2\mu)$.

Finally, we would like to relate the quantity $Z$ to the physical observables,
namely, to the scattering length $a$ and effective range $r$.
Here, we are closely following the path of Ref.~\cite{Weinberg}.
It is important to note that these relations can be derived when the binding
energy is much smaller than the inverse of the range of forces. We start with
the twice-subtracted dispersion relation for the inverse of $t(E)$
\begin{equation}
    t^{-1}(E) = \frac{E+E_B}{g_\text{NR}^2 } + \frac{(E+E_B)^2}{\pi}
\int_0^{\infty}\!\!dw \frac{\text{Im}\,t^{-1}(w) }{ (w-E-i\epsilon)(w+E_B)^2 },
\label{eq:dispersion}
\end{equation}
where the two subtraction constants have been determined from
Eq.~\eqref{eq:low}. The S-wave transition matrix element is related to the
non-relativistic S-wave scattering amplitude $f(k)=1/[k\cot\delta(k)-i\,k]$ as
$f(k) = -\mu\,t(E)/(2\pi)$ with $k=\sqrt{2\mu E}$ and $\delta(k)$ being
the S-wave phase shift. Thus, one gets
$\mbox{Im}\,t^{-1}(w)=\mu\sqrt{2\mu w}/(2\pi)$. Inserting this into
Eq.~\eqref{eq:dispersion}, we obtain
\begin{equation}
    t^{-1}(E) = \frac{E+E_B}{g_\text{NR}^2 } + \frac{\mu}{4\pi}R \left( \frac1R
+i\,k \right)^2,
\end{equation}
where $R=1/\sqrt{2\mu E_B}$ denotes the characteristic distance
between the constituents
in the two-body bound system.
Comparing the above expression with the effective
range expansion $t^{-1}(E) = - \mu/2\pi \left(-1/a + r\, k^2/2 - i\,k\right)$,
and using Eq.~\eqref{eq:gnr}, one can express the
scattering length and effective range in terms of the binding energy and
compositeness~\cite{Weinberg}
\begin{equation}
    a = \frac{2\,R\,(1-Z)}{2-Z}, \qquad r = -\frac{R\,Z}{1-Z}.
\end{equation}
Therefore, for an S-wave shallow two-body bound state, the compositeness can
be measured by measuring the low-energy scattering parameters.

Next, we turn to the compositeness condition within the framework of
the quantum field theory. For simplicity, let us first consider the situation
when a scalar particle described by a field $\Phi(x)$ with the bare mass
$M_0$ couples with two scalars
$\phi_{1,2}(x)$ with the masses $m_{1,2}$.
The interaction Lagrangian takes the form
${\cal L}_{\sf int}=g_0\, \Phi \phi_1\phi_2$.

\begin{figure}[t]
    \includegraphics[width=10cm]{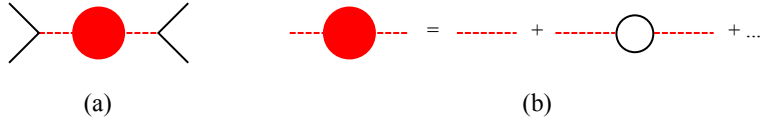}
\caption{The scattering matrix for the process
$\phi_1\phi_2\to\phi_1\phi_2$ (a) and
the two-point function of the field $\Phi$ (b). Only one-loop bubbles are summed
up. Solid (dashed) lines denote $\phi_{1,2}$ ($\Phi$) fields, respectively.}\label{fig:bubble}
\end{figure}

Consider now the two-point function of the field $\Phi(x)$
\begin{eqnarray}
    \mathcal{G}_\Phi(s)  = \int\! d^4 x\, e^{i\, Px} \left\langle
0 \left| T
\Phi(x)\,\Phi (0) \right| 0 \right\rangle \, ,\quad \text{with}
\quad s=P^2\, .
\end{eqnarray}
Summing up one-loop bubble diagrams to the two-point function, one
arrives at the expression (see Fig.~\ref{fig:bubble})
\begin{eqnarray}
    \mathcal{G}_\Phi(s) = \frac{i}{ s - M_0^2 - g_0^2\, G(s) }\, ,
\end{eqnarray}
where the one-loop self-energy is given by
\begin{equation}
 G(s) = i\int\frac{d^4q}{(2\pi)^4}\frac{1}{(P-q)^2-m_1^2+i\epsilon}
  \frac{1}{q^2-m_2^2+i\epsilon}.
  \label{eq:Gloop}
\end{equation}
The relativistic scattering amplitude for the process
$\phi_1\phi_2\to\phi_1\phi_2$ in the same approximation is given by (see
Fig.~\ref{fig:bubble})\footnote{Here, in order to be consistent with the non-relativistic formalism, the sign convention $S=1-iT$ is used in the definition of the $T$-matrix.}
\begin{eqnarray}\label{eq:Ts}
T(s)= \frac{g_0^2}{ s - M_0^2 - g_0^2\, G(s) }\, .
\end{eqnarray}
The relativistic and the non-relativistic scattering matrices are the same
up to an overall normalization. In the rest frame of the bound system, the
relation takes the form
\begin{eqnarray}
T(s)= 4w_1(k)w_2(k)\, t(E)\, ,\quad\quad E=\sqrt{s}-(m_1+m_2)\, ,
\end{eqnarray}
where $w_i(k) = \sqrt{m_i^2+k^2}$.
Now, let us consider the behavior of the scattering amplitude in the vicinity
of the bound-state pole. The two-point function has the following behavior
\begin{eqnarray}
    \mathcal{G}_\Phi(s) \to
  = \frac{i\, Z} {s - M^2 + i\epsilon} + \mbox{less singular terms}\, ,
\quad M^2=M_0^2+g_0^2G(M^2)\, ,
\end{eqnarray}
where $M$ is the physical mass.

The residue of the propagator determines the wave function renormalization
constant for the particle $\Phi$:
\begin{equation}\label{eq:ZRel}
    Z = \frac1{  1 - g_0^2\,G'(M^2) } = 1 +  g^2\,G'(M^2),
\end{equation}
where $g^2=Z\,g_0^2$ is the renormalized coupling constant, and 
$G'(M^2) = \left.\frac{d}{ds}G(s) \right|_{s=M^2} $.
In order to establish the relation of the quantity $Z$, defined by
Eq.~(\ref{eq:ZRel}), with its non-relativistic counterpart, we
 perform the contour integration over $q^0$ of the
loop integral in Eq.~\eqref{eq:Gloop}:
\begin{equation}\label{eq:Gloop3}
    G(s)  =  \int\!\frac{d^3\vec q}{(2\pi)^3}\frac1{2\omega_1\omega_2}
  \frac{\omega_1+\omega_2}{s+\vec P^2-(\omega_1+\omega_2)^2+i\epsilon},
\end{equation}
where $\omega_1^2 = (\vec P-\vec{q}\,)^2+m_1^2$ and $\omega_2^2 =
\vec{q}^{\,2}+m_2^2$. In the rest frame of the bound state, one has $\vec P =
0$. Taking derivative with respect to $s$, and then taking the non-relativistic
approximation which amounts to $\omega_1\simeq m_1 +
\vec{q}^{\,2}/(2m_1)$ and $\omega_2\simeq m_2 + \vec{q}^{\,2}/(2m_2)$, we get
\begin{equation}
    g^2\,G'(M^2) \simeq  - \frac{g^2}{8m_1 m_2 M}
\int\!\frac{d^3\vec q}{(2\pi)^3}
\frac1{ \left[ E_B + \vec{q}^{\,2}/(2\mu) \right]^2 },
\end{equation}
where we have used $E_B = m_1+m_2 - M$. Taking into account the difference
between relativistic and
non-relativistic normalizations, we finally arrive at the relation $g =
\sqrt{2m_1}\sqrt{2m_2}\sqrt{2M} g_\text{NR}$, cf. with Eq.~(\ref{eq:low}).
Comparing now this relation with
Eq.~\eqref{eq:BB0}, one immediately sees that the wave function renormalization
constant $Z$ is the same as its non-relativistic counterpart and thus the
compositeness condition for an S-wave bound state can be written as
\begin{equation}
  \label{1-Z}
  Z =1+g^2\,G'(M^2)\to 0.
\end{equation}
One might treat the above argumentation with a grain of salt, since it is
based on certain approximations. Namely, the amplitude is given as a sum
of one-loop diagrams only. It is, however, clear that the result is valid
beyond this approximation, if bound states close to an elastic threshold
are considered. The justification is provided by the statement that such bound
states can be consistently described within a non-relativistic effective
field theory, which is perturbatively matched to the underlying relativistic
theory (see, e.g., Ref.~\cite{Gasser:2007zt} for a review on the subject).
Such an effective theory is equivalent to the non-relativistic
quantum mechanics (the number of particles is conserved) and hence the
compositeness can be rigorously defined along the lines discussed above.
Finally, we would like to mention that the quantity $Z$, which is defined
in Eq.~(\ref{eq:ZRel}), is ultraviolet finite, since the quantity $g$
is defined through the residue of the {\it renormalized} scattering amplitude.

\section{Compositeness from lattice data}
\label{sec:mass-shift}

As stated above, the wave function renormalization constant, $Z$, gives an
overlap of the physical state with the elementary state and hence could be used
as a parameter that describes the compositeness of a given state.
Lattice calculations provide a model-independent way to determine $Z$ from the 
volume dependence of the
spectrum~\cite{Sekihara:2012xp,Luscher:1985dn,Beane:2003da,Koma:2004wz,
Sasaki:2006jn,Davoudi:2011md}, or -- as we propose in this paper -- from the dependence
on the twisting angle. In this section we set up
a finite-volume formalism, which describes the dependence of the bound-state mass on the volume
or twisting angle.

\subsection{Finite volume formalism}

We consider elastic scattering of particles with
the masses $m_1$ and $m_2$ in the S-wave\footnote{In order to make the
presentation transparent, throughout this paper we do not consider the 
partial-wave mixing in a finite volume. This effect can be 
later included in a standard manner.}. Then, generally,
a unitary partial-wave amplitude in infinite volume is given by
\begin{equation}
  \label{Tamplitude}
  T(s) = \frac{1}{V^{-1}(s)-G(s)} = \frac{-8\pi\sqrt{s}}{k\cot\delta(k)-ik},
\end{equation}
where $k^2=\frac1{4s}[s-(m_1+m_2)^2][s-(m_1-m_2)^2]$
is the relative momentum squared in the center of mass (c.m.) frame.
Further, the function $V^{-1}(s)$ (``the inverse potential'')
is a regular function in the vicinity of the
threshold.
The notation used here is
reminiscent of that of unitarized Chiral Perturbation Theory, but
Eq.~\eqref{Tamplitude}
may in fact describe any elastic unitary amplitude, with the particular dynamics encoded
in the
function $V(s)$. The loop function $G(s)$ is given by Eqs.~\eqref{eq:Gloop} and \eqref{eq:Gloop3}.
This function contains a unitarity cut. Across this cut,
we have $\im G(s) = -k / (8\pi\sqrt{s})$. Other (distant) 
cuts that may be also present are included in $V(s)$.
The loop function $G(s)$ is divergent and has to be renormalized. Here we do the renormalization
with a subtraction constant. As it will be seen below, the extension to the finite volume is
independent of any regulator.

When the particles are put in a finite box of size $L$, their momenta
become discretized due to boundary conditions.
So, the continuum spectrum, which gives rise to the cut in the infinite volume, becomes a discrete
set of two-particle levels. In order to obtain the spectrum in a finite volume,
one should replace
the momentum integrals by the sums over the discretized momenta in the expression of the scattering amplitude.
Then, the ``finite volume scattering amplitude'' $\tilde T$ contains 
poles on the real axis that correspond to the discrete two-particle levels. 
It should be noted that the finite-volume effects in $V(s)$ are exponentially
suppressed (see, e.g.,~\cite{luescher-torus}), so the 
the finite volume scattering amplitude can be obtained just by changing the loop
function by its
finite volume counterpart $\tilde G_L^{\vec\theta}(s) = G(s)+\Delta
G_L^{\vec\theta}(s)$~\cite{Doring:2011vk}, where
\begin{equation}
  \label{deltaG}
  \Delta G_L^{\vec\theta}(s) = \lim_{\Lambda\to\infty}
  \left[
    \frac1{L^3}\sum_{|\vec q_n|<\Lambda} I(\vec q_n) -
    \int_{|\vec q|<\Lambda}\frac{d^3\vec q}{(2\pi)^3}I(\vec q\,)
  \right]\, .
\end{equation}
Here $I(\vec q\,)$ denotes the integrand in  Eq.~\eqref{eq:Gloop3},
and $\vec q_n$
the allowed momenta in a finite volume, whose value
depends on the box size $L$ and the
boundary conditions used. For the periodic boundary conditions
we have $\vec q_n=\frac{2\pi}{L}\vec n$, $\vec n\in\mathbb{Z}^3$. In case of
twisted boundary conditions, the momenta also depend on the twisting angle
$\vec\theta$ according to $\vec q_n=\frac{2\pi}{L}\vec n+\frac{\vec\theta}{L},~0\leq\theta_i<2\pi$.
Using the methods of Ref.~\cite{Doring:2011vk}, it can be shown that $\Delta G_L^{\vec\theta}$ can be
related to the modified L\"uscher function $Z_{00}^{\vec\theta}$, see Appendix~\ref{appendix-deltaG},
\begin{equation}
  \label{deltaG-luscher1}
  \Delta G_L^{\vec\theta}(s)=\frac1{8\pi\sqrt{s}}\left(
    ik - \frac{2}{\sqrt{\pi}L} Z_{00}^{\vec\theta}(1,\hat k^2)
  \right)+\cdots,
\end{equation}
where $\hat k = kL/(2\pi)$ and the dots stand for terms that are exponentially
suppressed with the volume size $L$~\cite{Doring:2011vk}.

In this paper, we are going to apply L\"uscher formalism to study
shallow bound states, where the finite-volume effects are exponentially 
suppressed. Since, for such states, the binding momentum $\kappa$ 
is presumed to be much smaller than the lightest mass in the system, the 
exponentially suppressed corrections emerging, e.g., from the potential
 $V(s)$ could be consistently neglected as compared to the corrections 
$\sim e^{-\kappa L}$ that arise from $Z_{00}^{\vec\theta}(1,\hat k^2)$. 
Note however
that, if masses of the constituents increase for a fixed binding energy, then
the magnitude of the binding momentum also increases and, for the bound states
of heavy mesons, may become comparable to the pion mass. In this case, further
study of the problem is necessary. A recent example of such a study (albeit 
in the light quark sector) is given in Ref.~\cite{Albaladejo}. 
In the present paper this issue is not addressed.

Finally, note that the divergences arising at $\Lambda\to\infty$ in
Eq.~\eqref{deltaG}
cancel between the sum and the integral, so we can safely send the cutoff
to infinity. Thus, $\Delta G_L^{\vec\theta}$ does not depend on any regulator.
In Appendix~\ref{appendix-deltaG} we show in detail, how $\Delta
G_L^{\vec\theta}$ could be calculated below threshold for different types of boundary conditions. 

\subsection{Bound states in finite volume}

Bound states show up in the scattering amplitude as poles on the real axis below
threshold.
Namely, if we have a bound state with the mass $M$ in the infinite volume,
the scattering amplitude should have a pole at $s=M^2$, with the corresponding binding momentum
$k_B\equiv i\kappa$, $\kappa>0$.
From Eq.~\eqref{Tamplitude}, it is clear that $M$ and $k_B$
satisfy the equation
\begin{equation}
  \label{inf-vol-pole-eq}
  \psi(k_B^2)+\kappa = -8\pi M\Big[V^{-1}(M^2) - G(M^2)\Big] = 0,
\end{equation}
where $\psi(k^2)$ is the analytic continuation of $k\cot\delta(k)$ for
arbitrary complex values of $k^2$, which is needed since the bound state
is located below threshold, $k_B^2<0$.
On the other hand, the discrete levels in a finite volume are obtained as the
poles of the
finite-volume scattering amplitude $\tilde T$ and, in particular, the
bound state pole gets shifted to $M_L$, with binding momentum $k_L\equiv
i\kappa_L$, given by
\begin{equation}
  \label{finite-levels}
  \tilde T^{-1}(M_L^2)=T^{-1}(M_L^2)-\Delta G_L^{\vec\theta}(M_L^2) =
0\quad\Rightarrow\quad
  \psi(k_L^2)+\kappa_L+8\pi M_L\Delta G_L^{\vec\theta}(M_L^2)=0\, .
\end{equation}
Note that, below threshold, both $T^{-1}$ and $\Delta G_L^{\vec\theta}$ are real,
so the pole position
 is real. The discrete scattering levels above threshold are real as well
(as they should be), since the imaginary part of $\Delta G_L^{\vec\theta}$
cancels exactly with that of $T^{-1}$.

Next, we relate the finite-volume pole position with the infinite-volume
quantities as the bound state mass, $M$, and  the coupling, $g^2$ (defined as
the residue of the scattering amplitude at the pole $s=M^2$). To this end,
we expand $\psi(k_L^2)$ around the infinite-volume pole position,
$k_B=i\kappa$,
\begin{equation}
  \label{psi-expand}
  \psi(k_L^2) \simeq \psi(k_B^2)-\psi'(k_B^2)(\kappa_L^2-\kappa^2) =
  -\kappa-\psi'(k_B^2)(\kappa_L-\kappa)(\kappa_L+\kappa),
\end{equation}
where the prime denotes a derivative respect to $k^2$.
Then, evaluating the residue at $M^2$ in Eq.~\eqref{Tamplitude} we obtain
\begin{equation}
  \label{psi'}
  \psi'(k_B^2) = \frac1{2\kappa}-\frac{8\pi M}{g^2\frac{dk^2}{ds}},
\end{equation}
where the derivative $dk^2/ds$ is to be evaluated at $s=M^2$.
 Finally, using Eqs.~\eqref{finite-levels} and \eqref{psi-expand},
we obtain for the pole position shift
\begin{equation}
  \label{mass-shift}
\kappa_L - \kappa = \frac1{1-2\kappa\psi'(k_B^2)} \left[ -8\pi M_L \Delta
G_L^{\vec{\theta}} (M_L^2) + \psi'(k_B^2)(\kappa_L-\kappa)^2 \right]
\end{equation}
This equation gives the bound state pole position, $\kappa_L$ (or,
equivalently,
$M_L=\sqrt{m_1^2-\kappa_L^2}+\sqrt{m_2^2-\kappa_L^2}$)  as a function of the
infinite-volume parameters $g^2$ and $\kappa$.
It is worth noting that, within the approximation~\eqref{psi-expand},
the position of the bound state pole in a finite volume depends only
on these two parameters. This
 approximation works remarkably well in all cases considered in this paper.

If the difference $\kappa_L-\kappa$ is small enough, Eq.~\eqref{mass-shift} can
be solved
iteratively. For periodic boundary conditions, with the use of Eq.~\eqref{deltaGpoiss},
it can be shown that the lowest-order iterative solution reads
\begin{equation}
  \label{kL-iteration}
  \kappa_L = \kappa+\frac{6}{1-2\kappa\psi'(k_B^2)}\,\frac1{L}\,e^{-\kappa L}\, ,
\end{equation}
which coincides with the result given in
Refs.~\cite{Sasaki:2006jn,Beane:2003da,Sekihara:2012xp}.
However, it will be shown  below that, 
for shallow bound states, where $\kappa$ is very small, one should take more
than
just the first term in the sum~\eqref{deltaGpoiss}.
Moreover, in some cases, the iterations converge very slowly, if at all.
Therefore, in our opinion, it is safer to consider
solving Eq.~\eqref{mass-shift} numerically, without further approximations, in order to
obtain the finite volume pole position $\kappa_L$. This is the way we proceed.

Using Eq.~\eqref{mass-shift}, it is possible to fit the
infinite-volume parameters $M$ and $g^2$ from the bound state levels $\kappa_L$,
obtained through lattice simulations at different $L$ or $\vec\theta$. 
This, in turn, allows one
to determine the compositeness parameter from Eq.~\eqref{1-Z}. However, in actual lattice
simulations, the measured energy levels have some uncertainty, and the number of different volumes or
different twisting angles might be not very large. Therefore,
it is important to know in advance, at which accuracy should be the lattice
measurements carried out, in order to render the extraction of the parameter $Z$ reliable.
We address this question in some exactly solvable models with a given $V(s)$, producing
``synthetic lattice data,'' adding random errors and trying to extract back the
 infinite volume parameters $M,g^2$ and $Z$ from data.

\section{Analysis with two  models}
\label{models}

\subsection{A toy model}
\label{toy-model}

The potential in this model is given by a ``bare state pole'',
\begin{equation}
  \label{Vtoy}
  V_{\sf toy}(s)=\frac{g_0^2}{s-s_0},
\end{equation}
which depends on two parameters: a bare pole position $s_0$ and a bare coupling
constant $g_0$.
By appropriately choosing the value of the bare parameters, we can reproduce 
a bound state with any given mass $M$ and coupling $g$.

If our model describes the interaction of two particles,
where a bound state with the mass $M$ is present,
the scattering partial wave amplitude~\eqref{Tamplitude} should have
a pole at $s=M^2$,
\begin{equation}
  \label{toy-pole}
  M^2-s_0-g_0^2G(M^2) = 0.
\end{equation}
The physical coupling of the bound state, $g$, is given by the
residue of the scattering partial-wave amplitude
at the bound state pole
\begin{equation}
  \label{toy-coupling}
  g^2=\frac{g_0^2}{1-g_0^2G'(M^2)}=[1+g^2G'(M^2)]g_0^2=Zg_0^2\, .
\end{equation}
One can use above equations to trade the bare parameters for the
physical ones in
the expression of the scattering amplitude and write the latter in terms of
$M$ and $Z$:
\begin{equation}
  \label{Ttoy}
  T_{\sf toy}(s)=\frac{Z-1}{(s-M^2)ZG'(M^2)+(1-Z)[G(s)-G(M^2)]}.
\end{equation}
Note that the above amplitude does not depend on the subtraction constant
that renders $G(s)$ finite.
This model can describe a bound state with any
given value of the wave function renormalization constant.

Next, we study the
finite volume effects in the bound-state mass.
In the actual calculations, we take
$m_1=m_D$, $m_2=m_K$ and
choose the mass of the bound state to be $M=2340$ MeV. This is a shallow
bound state at 20~MeV below
threshold, which corresponds to a binding momentum
$\kappa\simeq 133$ MeV.
For the mainly molecular state we take $Z=0.1$, and
$Z=0.9$ is chosen for the mainly elementary one.
For each of these two states, we
calculate their finite-volume mass $M_L$ as the subthreshold pole position in
the finite-volume scattering amplitude.

\begin{figure}[t]
  \centering
  \hbox{
    \includegraphics[scale=.33, angle=-90]{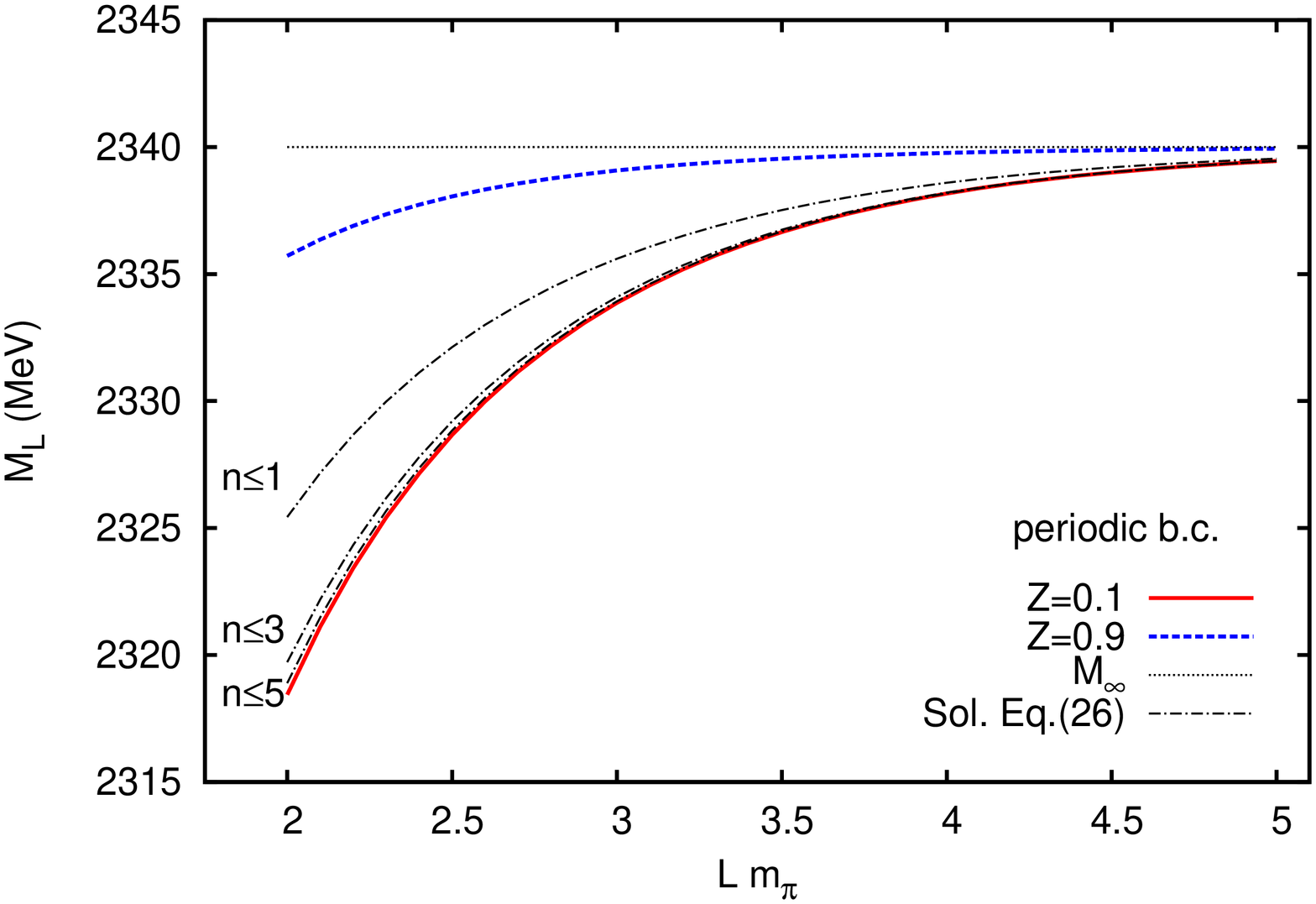}
    \includegraphics[scale=.33, angle=-90]{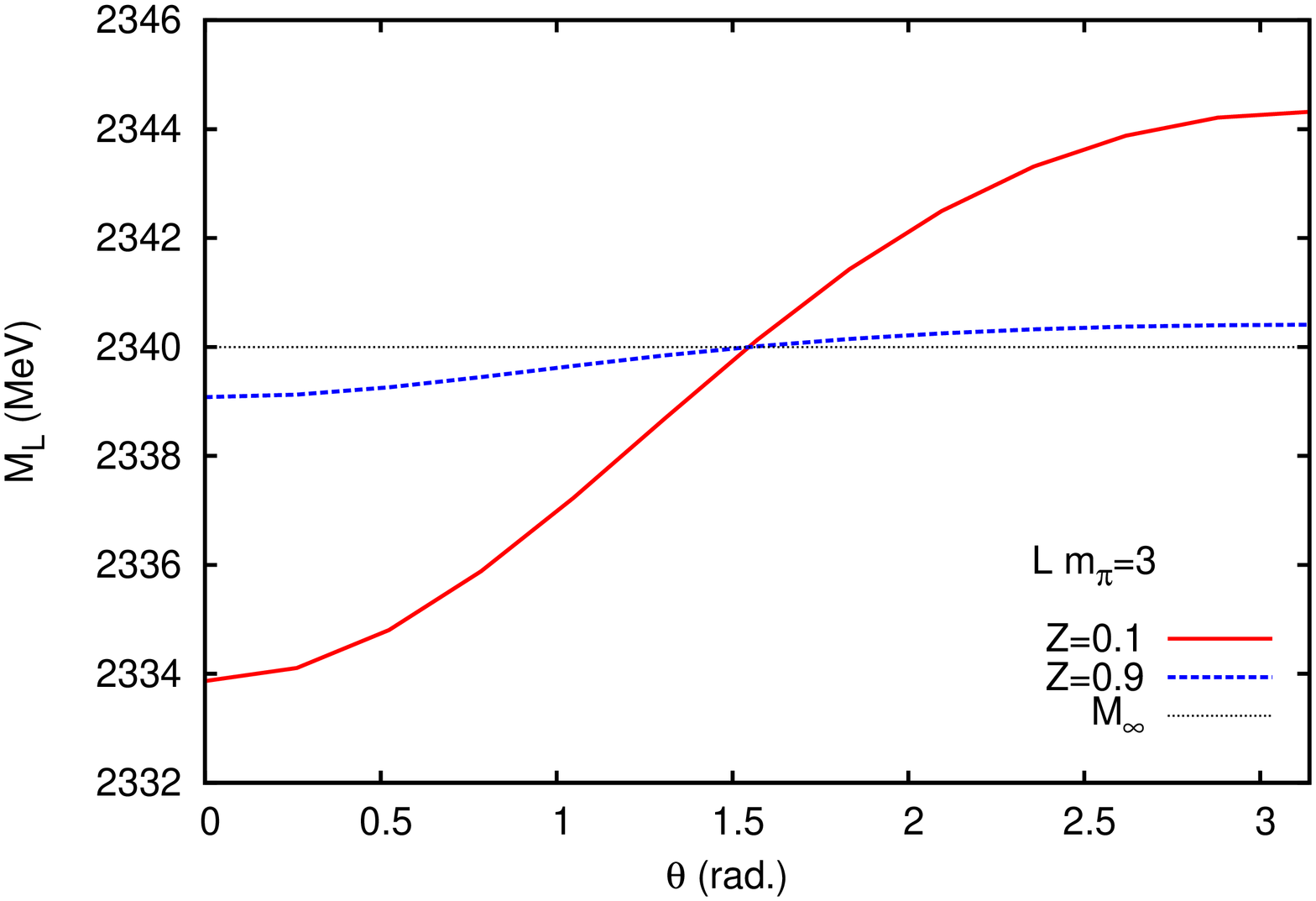}
  }
  \caption{Bound state mass in the finite volume, $M_L$, as a
    function of $L$ for periodic boundary conditions (left)
    and as a function of the twisting angle for twisted boundary conditions
(right).
    The solid/dashed lines correspond to $Z=0.1$ and $Z=0.9$, respectively. 
The dotted line stands for the infinite-volume mass $M$.
In order to test the accuracy of the iterative solution,
    for the case of $Z=0.1$ we also plot (dot-dashed lines) the solutions of 
Eq.~\eqref{mass-shift} with an approximate expression of $\Delta G_L^{\vec\theta}$
(only the first $n\equiv |\vec n|$ terms are retained in the expression
\eqref{deltaGpoiss} for $\Delta G_L^{\vec\theta}$).}
  \label{toy-dependence}
\end{figure}

In the left panel of Fig.~\ref{toy-dependence}, we show the mass
of the two states with $Z=0.1$ and $Z=0.9$
as a function of $L$ for periodic boundary conditions\footnote{Note that throughout this paper we take the physical value of $m_\pi$ and do not discuss the pion mass dependence.}. These are obtained 
from the solution of the exact equation~\eqref{finite-levels}.
It is easy to see that the finite volume effects are much bigger in
the case of the
molecular state with $Z=0.1$ than in the case of an elementary state with
$Z=0.9$.
This was of course expected in advance, since small finite-volume effects point 
on a compact nature of the state in question.
Here we also plot the solutions of Eq.~\eqref{mass-shift},
 using the known values of
$M$ and $g$, taken from the infinite volume model. In this way we can test the
validity of the
approximation in Eq.~\eqref{psi-expand}, used to derive
Eq.~\eqref{mass-shift} from Eq.~\eqref{finite-levels}, which
basically states that all relevant dynamics is encoded only in the two
parameters $M$ and $g$.
As can be seen in Fig.~\ref{toy-dependence}, Eq.~\eqref{mass-shift} is able to
reproduce
the synthetic lattice results very accurately.
On the other hand, note that for shallow bound states the binding momentum
$\kappa$ is small, so no wonder that
the expansion in $\Delta G_L^{\vec\theta}$ converges 
rather slowly. Consequently, retaining only the leading-order term and 
constructing iterative solution, see Eq.~\eqref{kL-iteration},
 might not be sufficient in all cases.

In the right panel of
the same figure we show the dependence of the bound-state mass 
on the twisting angle $\vec\theta=(\theta,\theta,\theta)$ for 
the fixed value of $Lm_\pi=3$.
We see that, for such a choice of twisting, the
size of the effect of twisting for a fixed $L$ is almost the double of
the maximal effect caused by the variation of $L$ from the same value 
to infinity (periodic boundary conditions).
Thus, using (partially) twisted boundary conditions to
determine $Z$, besides being cheaper,
could give more accurate results than a method based on the study
of the volume-dependence of the energy level.
Note also that, for the above choice of the twisting angle, the twisting
effect is maximal. Other choices, e.g., $\vec\theta=(0,0,\theta)$ lead to a 
smaller effect.

\subsection{$DK$ scattering and the $D_{s0}^*(2317)$}
\label{DK}

Now we turn our attention to the realistic case
of the hadronic bound state $D_{s0}^*(2317)$ in
the $DK$ scattering channel with isospin $I=0$ and
strangeness $S=1$. When isospin symmetry is exact, this state is stable
under strong interactions, since it does not couple to the 
lighter hadronic channels
(the observed decay $D_{s0}^*(2317)\to D_s\pi$ breaks isospin symmetry).
Thus, the formalism above, tailored for stable bound states,
does apply in this case. The case of quasi-bound states, which are
coupled to inelastic channels, requires special treatment and is not 
addressed here.

A popular view on the $D_{s0}^*(2317)$ meson is that this state is dynamically
generated as a pole through the S-wave interactions
 between the $D$-meson and the kaon in the isoscalar
channel~\cite{Kolomeitsev:2003ac,Guo:2006fu,Hofmann:2003je,Gamermann:2006nm,
Guo:2009ct,Liu:2012zya}.  We shall study this
system, using the model used from Ref.~\cite{Guo:2006fu}, 
which is based on the leading-order heavy flavor chiral 
Lagrangian~\cite{Burdman,Wise:1992hn,Yan:1992gz}
and unitarizes the 
amplitude~\cite{Oller:1997ti,Oller:1997ng,Oller:1998hw,Nieves:1998hp}.
Namely, the 
infinite-volume amplitude is obtained from Eq.~\eqref{Tamplitude} with
the S-wave-projected potential
\begin{equation}
  \label{DKpotential}
  V(s)=\frac{1}{2}\int_{-1}^{1}dx\,\frac{u(s,x)-s}{2f_\pi^2}=
  \frac1{2f_\pi^2}
  \left[
    m_D^2+m_K^2+\frac{(m_D^2-m_K^2)^2}{2s}-\frac{3s}{2}
  \right],
\end{equation}
where $x=\cos\theta$ is the cosine of the scattering angle, $f_\pi\simeq 92.4$
MeV is
the pion decay constant, and $s$ and $u$ are usual Mandelstam
variables. We regularize the loop function with a subtraction constant 
$a(\mu)$, as done in Refs.~\cite{Oller-Meissner,Guo:2006fu}. Its value at the scale
$\mu=m_D$
is taken to be $a(m_D)=-0.71$. 
With this value of the subtraction constant, we find
a bound state pole, associated with the $D_{s0}^*(2317)$, at 
$M=2316.9~\mbox{MeV}$,
and the coupling to $DK$, which is given by the residue of the pole,
\begin{equation}
  \label{g2def}
  g^2=\lim_{s\to M^2}(s-M^2)T(s),
\end{equation}
takes the value $g=10.7~\mbox{GeV}$. 
One can easily calculate the compositeness parameter of
the bound
state as well,
using Eq.~\eqref{1-Z}. The calculation yields $Z=0.29$. Hence, in this
model, the $D_s^*(2317)$ is predominately a molecular state.

Next, we study this model in a finite volume and
 consider twisting of different quarks, from which
the $D$ and $K$ mesons consist. The net effect is that these mesons get 
different momenta as a result of such twisting, so the expression
for $G_L^{\vec\theta}$ changes. Note that this issue is important in view
of the fact that partial twisting is allowed only for certain quarks
(see Section~\ref{partially} for more details).

\begin{figure}[t]
  \centering
  \hbox{
 \includegraphics[angle=-90,scale=0.33]{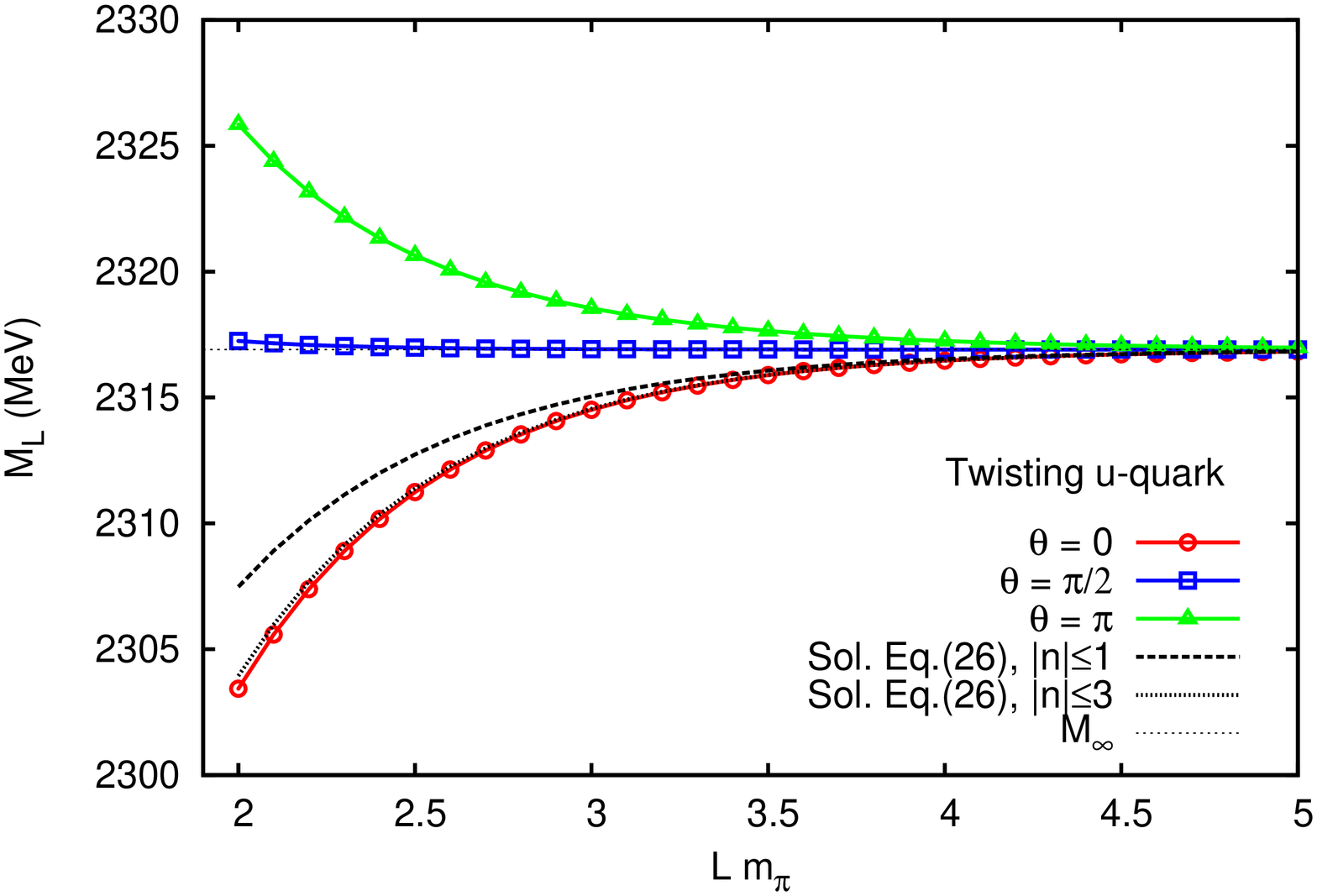}
 \includegraphics[angle=-90,scale=0.33]{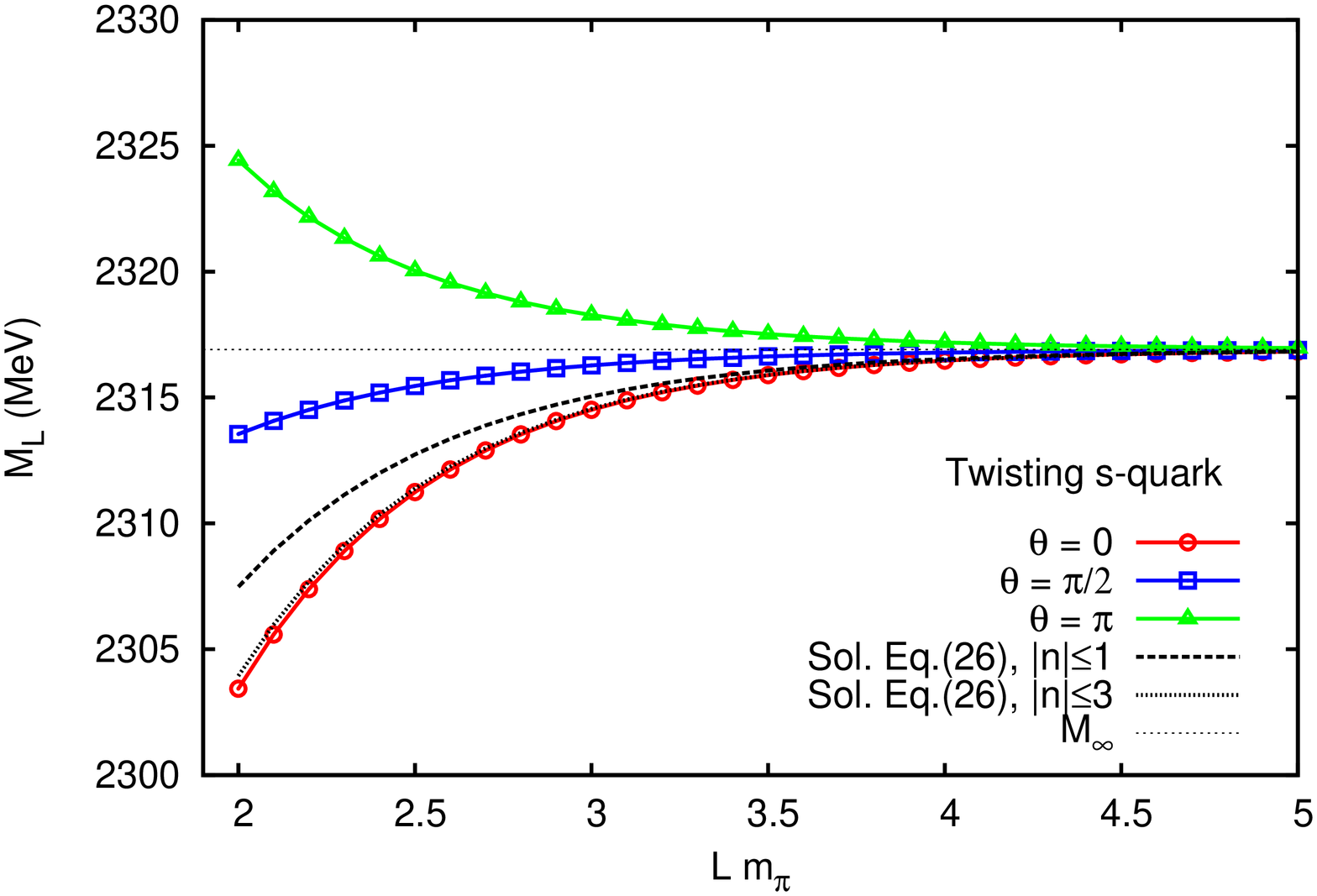}
  }
  \caption{$L$-dependence of the $DK$ bound-state mass for different twisting
angles. Left: twisted boundary
conditions applied to the $u$-quark. Right: twisted boundary conditions applied
to the $s$-quark.
The dashed lines give the solution of Eq.~\eqref{mass-shift},
using the values for $M$ and $g$
from the infinite-volume model. In these solutions, approximate expression   
for $G_L^{\vec\theta}$ at $\vec\theta=0$ was used, that amounts to summing
up exponentials only up to $|\vec n|\le n_{max}$.}
  \label{DK-L-dependence}
\end{figure}

In Fig.~\ref{DK-L-dependence}, we display
 the volume dependence of the bound state mass
for different twisting angles which are again chosen as
$\vec\theta=(\theta,\theta,\theta)$.
In the left panel, we plot the $L$-dependence
for three different values of the twisting angle, when twisted boundary
conditions are applied to the $u$-quark. In the right panel, twisted
boundary conditions are applied to the
$s$-quark. As we shall see later, in the latter case the 
use of partial twisting gives the same results as using fully twisted
boundary conditions. The size of the finite volume
effects, using twisted boundary conditions for the $c$-quark,
is very small, so we do not discuss this case.  In this model, 
we test again that the
predictions obtained from Eq.~\eqref{mass-shift}, using the values
of $M$ and $g$ from the infinite-volume model, reproduce very well the exact
solution. Consequently, all relevant dynamics of the model near threshold is 
 encoded in just two parameters $g$ and $M$.
On the other hand, we see that retaining only the leading exponential
in the expansion of $G_L^{\vec\theta}$ will have a large impact on the accuracy.
Consequently, the first few terms should be retained. We see that the 
convergence is satisfactory: e.g., taking $n_{max}\ge 3$, where $n_{max}$
denotes the number of terms retained in the expansion, we see that the
largest difference
between the synthetic data and the prediction from Eq.~\eqref{mass-shift} is
less than
0.1~MeV.

Analyzing Fig.~\ref{DK-L-dependence}, we again come to the conclusion
 that the use of (partially) twisted boundary
conditions can provide a better way to
extract the compositeness parameter $Z$ 
from lattice results.
This can already be seen by
comparing the curves for $\theta=0$ and $\theta=\pi$. One namely observes that 
the size of the effect due to twisting at a fixed volume
is almost twice as big as due to changing the volume for periodic boundary
conditions.

\begin{figure}[t]
  \centering
  \hbox{
    \includegraphics[angle=-90,scale=0.33]{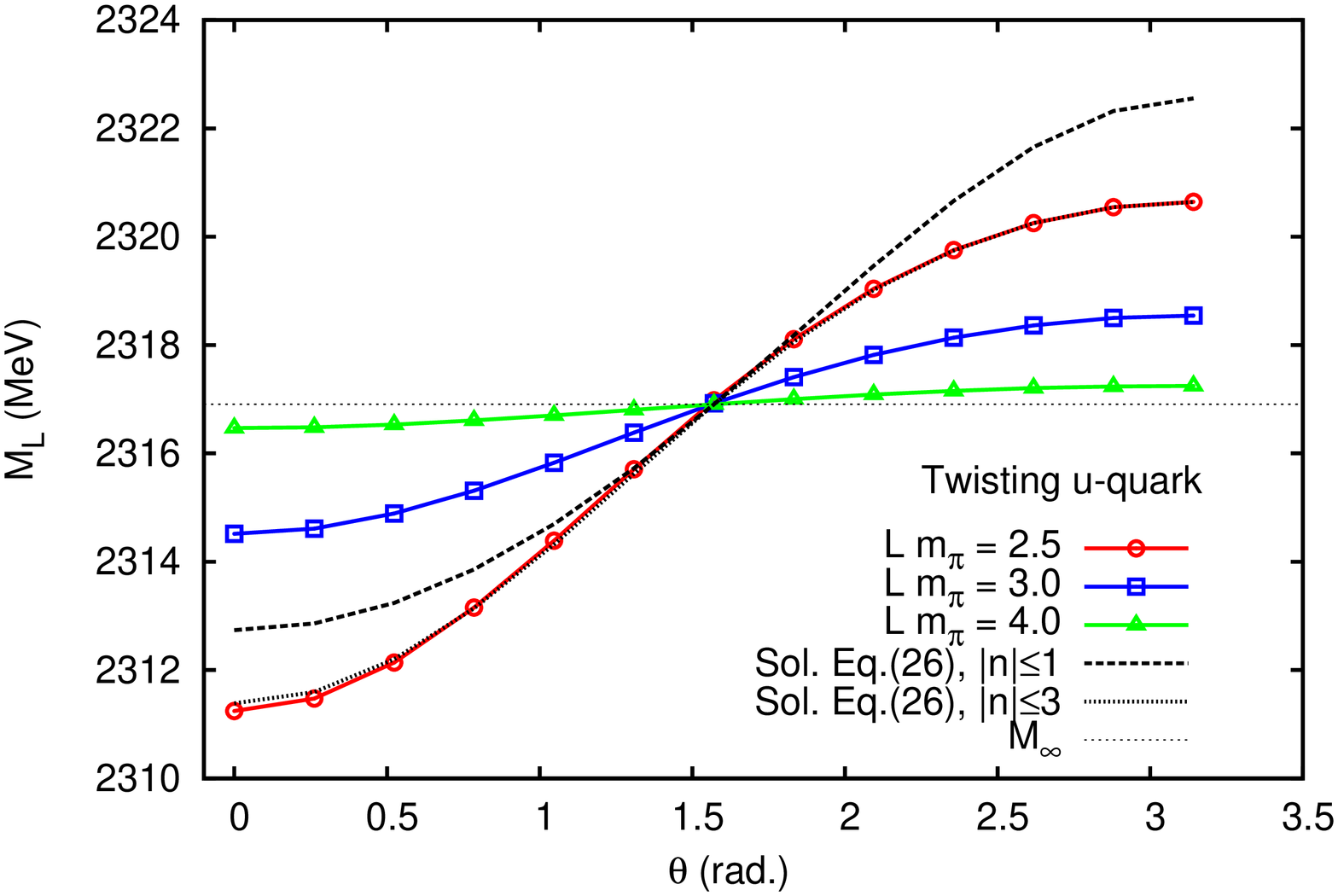}
   \includegraphics[angle=-90,scale=0.33]{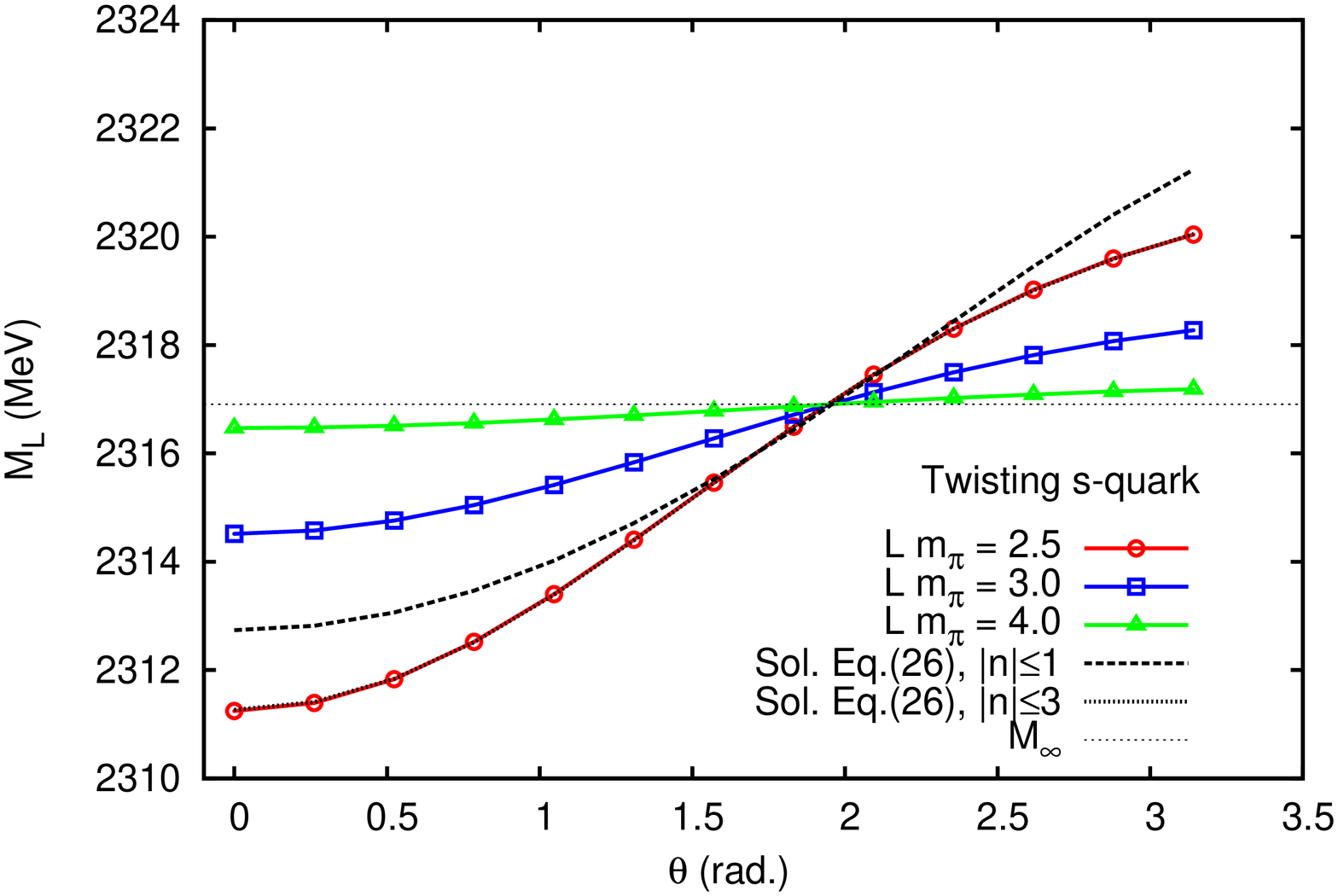}
  }
  \caption{$\theta$ dependence of the $DK$ bound state mass for
    different lattice volumes. Left: twisted boundary
conditions applied to the $u$-quark. Right: twisted boundary conditions applied
to the $s$-quark.
The dashed lines give the solution of Eq.~\eqref{mass-shift},
using the values for $M$ and $g$
from the infinite-volume model. In these solutions, approximate expression   
for $G_L^{\vec\theta}$ at $\vec\theta=0$ was used, that amounts to summing
up exponentials only up to $|\vec n|\le n_{max}$.}
  \label{DK-theta-dependence}
\end{figure}
In Fig.~\ref{DK-theta-dependence}, for three different volumes, we show the
dependence of the bound-state mass on the twisting angle both 
for $u$- and $s$-quark twisting. On the other hand, taking the
results of the $\theta$-dependence 
(like in Fig.~\ref{DK-theta-dependence}) at a fixed volume for 
granted,
one could fit the value of the infinite-volume mass and coupling constant to
these data, using
Eq.~\eqref{mass-shift}. After this, it is straightforward to obtain the
value of $Z$. In fact, producing four synthetic lattice data points at a
fixed $Lm_\pi=2.5$ and $\theta=0,\pi/3,2\pi/3,\pi$ (either for $u$- or $s$-
quark twisting), we were able to obtain values for $M$ and $g$
that differ less than 1\% from those calculated from the infinite volume model
by
fitting the solution to Eq.~\eqref{mass-shift} (with $n_{max}=5$) to the
synthetic data.

Real lattice simulations, however, produce results which carry uncertainties.
Hence, the question arises, how big these errors could be in order to be 
still able to
determine $Z$ with a desired accuracy.
Since, as seen from the figures, the  finite
volume effects (for reasonable volume sizes, say, 
above $Lm_\pi=2.5$) are at most
around 10 MeV, one expects that a relatively high accuracy will be needed
in the measurement of the bound-state energy. In order to determine,
how high this accuracy should actually be, we
assign an uncertainty to the synthetic data that we generate from our model.
In particular, using the von Neumann rejection method, from the
``exact'' data points we generate a new, ``randomized'' data set,
where the central values of each data point are shifted randomly, following the
Gaussian
distribution centered at exact data values and 
with a standard deviation, given by
the lattice data error. Repeating this process several times, we obtain several
sets of synthetic lattice data with errors and central values shifted
accordingly.
We then fit each of the randomized data sets and obtain a corresponding value
for $M$ and $g$ (and therefore, for $Z$), one for each set, ending up with
as many values for the parameters, as many 
randomized data sets we have generated.
We can obtain then the mean and standard deviation of the distributions for $M$,
$g$ and $Z$.
Thus, for a given data error, we can estimate the accuracy of the
parameter extraction.

\begin{table}[t]
  \centering
  \begin{tabular}{|c|cc|cc|}
    \hline
     &\multicolumn{2}{c|}{4 lattice data points}
    & \multicolumn{2}{c|}{8 lattice data points} 
    \\\hline
    $\Delta M_L$ (MeV) & $\;Lm_\pi=2.5\;$ & $\;Lm_\pi=3.0\;$ & $\;Lm_\pi=2.5\;$ & $\;Lm_\pi=3.0\;$ \\\hline
    2   & 0.21 & 0.47 & 0.17 & 0.36 \\
    1   & 0.11 & 0.23 & 0.08 & 0.19  \\
    0.5 & 0.05 & 0.12 & 0.04 & 0.09  \\
    \hline
  \end{tabular}
  \caption{The accuracy of the extraction of the parameter $Z$
from the fits to the synthetic lattice data for different input
error $\Delta M_L$.
    Four or eight data points and two different volumes $Lm_\pi=2.5$ and
    $Lm_\pi=3.0$ were used, see main text for details.}
  \label{fits}
\end{table}
 
For the case of the $s$-quark twisting,
we construct 5000 sets of randomized data at a fixed volume,
for different input errors $\Delta M_L$ and
different number of data points per set.
Fitting the parameters to each set, we obtain the corresponding
distributions of 5000 points for each parameter $M$, $g$ and $Z$. In
table~\ref{fits}, we show the resulting standard deviations for $Z$, 
which give an idea of the expected accuracy in a
fit to actual lattice data. The results for
the case of the $u$-quark twisting are very similar. We see that, for
$Lm_\pi=2.5$
where the finite volume effects are the largest, we need lattice errors smaller
than 1 MeV  in order to obtain an accuracy in $Z$ below 0.1.
For larger volumes, the accuracy required in the input lattice data is even
bigger.
If we increase the number of lattice data points, we get slightly
better results but, in general, the dependence on the increase of the size
of the data set is very mild. 
For example, we need to use around 20 data points to achieve an accuracy
of order 0.1 in $Z$, given an input error $\Delta M_L=2~\mbox{MeV}$
and volume $Lm_\pi=2.5$.

\section{Partially twisted boundary conditions in the $DK$ system}
\label{partially}

The partial twisting, unlike the full twisting, is more 
affordable in terms of computational cost in lattice simulations, 
because one does not need to generate new gauge
configurations.
Thus, it is very interesting to study whether it is possible to extract any
physically relevant information from simulations using this kind of boundary
conditions.
Problems may arise when there are annihilation channels present, 
as is the case in the 
$DK$ scattering in the isoscalar channel, where light quarks may annihilate.
An analysis of L\"uscher approach with partial twisting for
scattering problem in the presence of annihilation channels was
recently addressed in~\cite{Agadjanov:2013kja}. Namely, a
modified partially twisted L\"uscher equation was derived for the $\pi\eta-K\bar
K$ coupled channel
scattering in the framework of non-relativistic EFT.

\begin{table}[t]
\centering
\begin{tabular}{|c|c|c|}
\hline
~Index~ & Channel & Quark content \\
\hline
1 & ~$|K_{\sf vv}D_{\sf vv}\rangle$~ &
~$-\frac{1}{\sqrt{2}}\,|u_{\sf v}\bar s_{\sf v}
c_{\sf v}\bar u_{\sf v}+d_{\sf v}\bar s_{\sf v}c_{\sf v}\bar d_{\sf v}\rangle$~\\[1.1mm]
2 & $|K_{\sf vs}D_{\sf vs}\rangle$ &
$-\frac{1}{\sqrt{2}}\,|u_{\sf s}\bar s_{\sf v}
c_{\sf v}\bar u_{\sf s}+d_{\sf s}\bar s_{\sf v}c_{\sf v}\bar d_{\sf s}\rangle$\\[1.1mm]
3 & $|K_{\sf vg}D_{\sf vg}\rangle$ &
$-\frac{1}{\sqrt{2}}\,|u_{\sf g}\bar s_{\sf v}
c_{\sf v}\bar u_{\sf g}+d_{\sf g}\bar s_{\sf v}c_{\sf v}\bar d_{\sf g}\rangle$\\[1.1mm]
\hline
\end{tabular}
\caption{Scattering channels for the case of $I=0$.}
\label{tab:channels}
\end{table}
Here, we address the same problem in the context of the $DK$ scattering.
The method is described in Ref.~\cite{Agadjanov:2013kja}, to which the reader is referred for further details.
Consider first the scattering in the infinite volume. We start from building the
channel space by tracking the
quarks of different  species following through the quark diagrams describing
the $DK$ scattering. It is clear that,
since only light  quarks may annihilate, the possible final states  contain
{\em valence}, {\em sea} or {\em ghost} light quarks with equal masses, as given
in table~\ref{tab:channels}.
Omitting channel indices, the resulting {\em algebraic}
 Lippmann-Schwinger equation couples 3 different channels
\eq\label{eq:LS-full}
T=V+VG_{DK}T\, ,
\en
 where $T$, $V$ and $G$ are given by $3\times 3$ matrices.

The free Green function is given by
\eq\label{eq:G}
G_{DK}(s) = G(s)~\mbox{diag}~(1,1,-1)
\en
where $G(s)$ is defined in Eqs.~\eqref{eq:Gloop} and \eqref{eq:Gloop3}, supplemented by the prescription that the integral is performed in dimensional regularization {\it after} expanding the integrand in powers of 3-momenta (see Refs.~\cite{Agadjanov:2013kja,cuspwe} for details).
The minus sign on the diagonal of the matrix $G$ arises due to fermionic nature
of $D$ and $K$ mesons
composed of valence and (commuting) ghost quarks.

\begin{figure}[t]
\begin{center}
\includegraphics[width=7.cm]{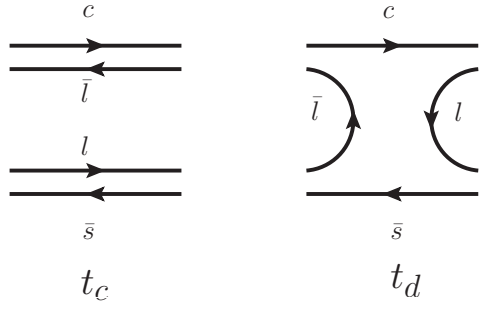}
\end{center}
\caption{Connected ($t_c$) and disconnected ($t_d$) diagrams, emerging in
  $DK\to DK$ scattering amplitudes with various quark species; $l$=$u,d$.}
\label{fig:cd}
\end{figure}

The crucial point now is that there exist linear symmetry relations between
various elements of $T$
due to equal valence, sea and ghost quark masses. Note that scattering
matrix elements
are given by residues of the 4-point Green functions $\Gamma_{ij}$ of the
bilinear quark operators at
 the poles, corresponding to the external mesonic legs. Decomposing
$\Gamma_{ij}$ into
 {\em connected}~ $t_c$  and {\em disconnected}~ $t_d$ pieces through Wick
contractions (see Fig.~\ref{fig:cd}) and
noting that quark propagators are the same for all light quark species, we get
\eq\label{eq:cd}
\hspace*{-.2cm}\Gamma_{11}=\Gamma_{22}=t_c-t_d\, ,\quad
\Gamma_{33}=-t_c-t_d\, ,\quad
\Gamma_{12}=\Gamma_{13}=\Gamma_{23}=\Gamma_{21}=\Gamma_{31}=\Gamma_{32}=-t_d\,
,
\en

Since in our case there are no neutral states and thus no mixing occurs, 
 following the
argumentation given in  Ref.~\cite{Agadjanov:2013kja}, it is easy to show that $T$-matrix obeys
the
same symmetry relations as $\Gamma$
 \eq\label{eq:ex1}
T_{11}=T_{22}=t,\quad
T_{33}=-t+2y\, ,\quad
T_{12}=T_{13}=T_{23}=T_{21}=T_{31}=T_{32}=y\, .
\en
Here $T_{11}=t$ corresponds to  the {\em physical} elastic $DK$ scattering
amplitude, i.e scattering
in the sector with valence quarks only. Other diagonal entries are {\em
unphysical} in
the  sense that they correspond to scattering of particles, composed of sea and
ghost light quarks.
Non-diagonal elements of $T$-matrix describe  coupling between valence and
sea/ghost sectors
through {\em disconnected}  diagrams.
Furthermore, it is straightforward to check from Eq.~(\ref{eq:LS-full}) that the
elements of
potential matrix $V$ satisfy the same symmetry relations as $T$ and can be
expressed in the
following form
\eq\label{eq:V-matrix}
V=\left(\begin{array}{c c c }
  \tau& \upsilon &\upsilon \\
 \upsilon& \tau&\upsilon \\
 \upsilon & \upsilon&-\tau+2\upsilon\\
\end{array}\right)\, ,
\en

Let us now  turn to the case of a finite volume and derive the L\"uscher equation for a couple of
particular
choices of partially twisted boundary conditions. Note that the potential $V$ remains
the same (up to
exponentially suppressed in terms $L$ ), while in the loop functions the integration
is
substituted by summation over lattice momenta. 

\begin{enumerate}
\item
Twist the $s$/$c$-quark, leaving  $u$ and $d$-quarks to obey periodic boundary
condition.  In this case, the matrix of the Green functions is
$\text{diag}\left( \tilde G^{\vec \theta}_L, \tilde G_L^{\vec 0}, -\tilde
G_L^{\vec 0} \right)$.
The solution of the Lippmann-Schwinger equation in a finite volume for the {\em physical} amplitude
$t$ is given by
\eq
t=\frac{\tau}{1-\tau\tilde G_L^{\vec\theta}}\, ,
\en
where $\tilde G_L^{\vec\theta}$ is the loop function $G(s)$ in a finite volume.
We see that the finite-volume spectrum in case of the partial twisting is determined
from the  L\"uscher equation
\eq
1-\tau\tilde G_L^{\vec\theta}(s)=0,
\en
in the same way as in the full-twisting case. Thus, the results obtained by using
of the partially
twisted boundary conditions on the $c$- or $s$-quark are equivalent to those
using full twisting.

\item
Twist the valence $u$-  and $d$-quarks simultaneously, leaving  $s$-
and $c$-quarks obey periodic boundary condition. In this case, the ghost
light quarks also need to be twisted, and the matrix of the Green functions is
$\text{diag}\left( \tilde G^{\vec \theta}_L, \tilde G_L^{\vec 0}, -\tilde G_L^{\vec \theta} \right)$.

The L\"uscher equation determining the finite volume spectrum now takes the form
\eq
\left[1-\tau\tilde G_{L}^{\vec 0}(s)\right]\left[1-(\tau-\upsilon)\tilde
G_L^{\vec\theta}(s) \right]^2=0\, .
\en
Vanishing of the first bracket on the r.h.s gives the L\"uscher equation with {\em
no twisting}. Note also that the quantity $\tau-\upsilon$ is in fact the connected part
of the scattering potential for the isoscalar $DK$ system, which is identical to
the $DK$ scattering potential in the isovector channel. Hence,
 vanishing of the second
bracket is equivalent to the fully twisted L\"uscher equation for the
isovector $DK$ scattering~\footnote{Since there is no disconnected Wick
contraction for the isovector $DK$ scattering, partial twisting is always
equivalent to the full twisting in this case.}.

\end{enumerate}
\section{Summary and conclusions}
\label{conclusions}

\begin{itemize}

\item[i)]
Lattice QCD does not only determine the hadron spectrum. Under certain
circumstances, it may provide information about the nature of hadrons, which
renders lattice simulations extremely useful for the search and the
identification of exotic states. Note that the lattice QCD possesses unique
tools at its disposal (e.g., the study of the volume and quark mass dependence
of the measured quantities), which are not available to experiment.

\item[ii)]
In the present paper, we concentrate on the identification of hadronic molecules
on the lattice. Experimentally, one may apply Weinberg's compositeness
condition to the near-threshold bound states, in order to distinguish the
molecular states from the elementary ones. To this end, one may use
the value of the wave function renormalization constant $Z$ which obeys
the inequalities $0\leq Z\leq 1$. The vanishing value of the parameter $Z$
corresponds to the purely molecular state. In this paper we consider the
lattice version of the Weinberg's condition.

\item[iii)]
It is known that the quantity $Z$ can be extracted from lattice data by
studying the volume dependence of the measured energy spectrum. We have
shown that the same result can be achieved by measuring the dependence of
the spectrum on the twisting angle in case of twisted boundary conditions.
Moreover, within the method proposed, the expected effect is approximately
twice as large in magnitude and comes at a lower computational cost.
Further, we have analyzed synthetic data to estimate the accuracy
of the energy level measurement which is required for a reliable extraction
of the value of $Z$ on the lattice.

\item[iv)]
As an illustration of the method, we consider the  $D_{s0}^*(2317)$ meson, which
is a candidate of a $DK$ molecular state. It is proven that, despite the
presence of the so-called annihilation diagrams, one may still use the
partially twisted boundary conditions for the extraction of $Z$ from data
if the charm or strange quark is twisted.
The effects which emerge due to partial twisting, are suppressed at large
volumes. 

\end{itemize}

The authors thank M. D\"oring, L. Liu, U.-G.~Mei{\ss}ner 
and S. Sasaki for interesting discussions.
This work is partly supported by the EU
Integrated Infrastructure Initiative HadronPhysics3 Project under Grant
Agreement no. 283286. We also acknowledge the support by the DFG (CRC 16,
``Subnuclear Structure of Matter''), by the DFG and NSFC
(CRC 110, ``Symmetries and the Emergence
of Structure in QCD''), by
the Shota Rustaveli National Science Foundation
(Project DI/13/02), by the Bonn-Cologne Graduate School of Physics and
Astronomy, by the NSFC (Grant No. 11165005), and by Volkswagenstiftung under
contract No. 86260.

\appendix
\section{Formulas for the function $\Delta G_L^{\vec\theta}$ below threshold}
\label{appendix-deltaG}

We compute the scattering amplitude in a finite volume by replacing the
loop function $G$ by its finite volume counterpart $\tilde
G_L^{\vec\theta}=G+\Delta G_L^{\vec\theta}$ 
and obtain synthetic data from
the poles of the finite volume scattering amplitude. In particular, the
pole below threshold
gives the mass of the bound state in a finite volume.

For the case of a level below threshold,
there exists a fairly simple way to calculate $\Delta G_L^{\vec\theta}$ defined
by Eq.~\eqref{deltaG},
 so that the equation~\eqref{mass-shift} for $\kappa_L$ can be easily solved. 
Here, we consider three different cases, 
one with periodic boundary conditions, and two with
twisted boundary conditions. Depending on which quarks are twisted,
the momenta of the mesons are modified accordingly. 

\subsection{Periodic boundary conditions}
In the case of periodic boundary conditions, the meson momenta in a box are
given by
\begin{equation}
  \label{allowed-q-periodic}
  \vec q_n=\frac{2\pi\vec n}{L},\qquad \vec n\in \mathbb{Z}^3.
\end{equation}
We can evaluate the sum in Eq.~\eqref{deltaG}, using the Poisson summation
formula
$\sum_n\delta(n-x)=\sum_n e^{2\pi i nx}$.
Transforming the sum into the integral gives
\begin{equation}
  \label{poisson}
  \frac1{L^3}\sum_{\vec n}I(\vec q_n)=\frac1{L^3}\sum_{\vec n}
  \int d^3\vec q\,\,\delta^{(3)}(\vec q-\vec q_n)I(\vec q\,)=
  \sum_{\vec n}\int\frac{d^3\vec q}{(2\pi)^3}\,\,e^{i\vec q\cdot\vec n L}
I(\vec q\, ).
\end{equation}
Next, we note that the integrand $I(\vec q\,)$ can be approximated by
$\frac{1}{2\sqrt{s}}\frac1{k^2-\vec q^2}$, since the difference is exponentially
suppressed~\cite{Doring:2011vk}.
Here, $k^2$ is the three-momentum squared of the
particles in the center of mass (c.m.) frame. Then, for $k^2<0$, 
$\Delta G_L^{\vec\theta}$
reads
\begin{eqnarray}
  \label{deltaGpoiss}
  \Delta G_L^{\vec 0}=\frac1{2\sqrt{s}}\sum_{\vec n\ne\vec 0}\int\frac{d^3\vec
q}{(2\pi)^3}\,
  \frac{e^{i\vec q\cdot\vec n L}}{k^2-\vec q^2}
  =-\frac{1}{8\pi\sqrt{s} L}
  \sum_{\vec n\ne\vec 0}\frac1{|\vec n|}e^{-|\vec n|\sqrt{-k^2}L}.
\end{eqnarray}
The function $\Delta G_L^{\vec 0}$ can be expressed in terms of the L\"uscher
zeta-function $Z_{00}(1,\hat k^2)$, as follows~\cite{Doring:2011vk}:
\begin{eqnarray}
  \label{deltaG-luscher}
  \Delta G_L^{\vec 0}&=&
  \frac1{8\pi\sqrt{s}}\left(-\sqrt{-k^2}-\frac{2}{\sqrt{\pi}L} Z_{00}(1,\hat
k^2)
  \right),\\[2mm]
  Z_{00}(1;\hat k^2)&=&\frac{1}{\sqrt{4\pi}}\,\sum_{{\vec n}\in \mathbb{Z}^3}
  \frac{1}{\vec n^2-\hat k^2}\, ,
\end{eqnarray}
where $\hat k = kL/(2\pi)$.

\subsection{Twisted boundary conditions: both momenta shifted}
In the case of twisted boundary conditions, when the momenta of both particles
are shifted but the particles still are in the c.m. frame, the allowed momenta in a box
are:
\begin{equation}
\label{allowed-q-twisted}
\vec q_n=\frac{2\pi}{L}\vec n+\frac{\vec\theta}{L},\qquad \vec n\in\mathbb{Z}^3\, ,
\end{equation}
where $\vec\theta$ is the twisting angle.
Now, acting in the same way, we can evaluate the sum in
Eq.~\eqref{deltaG}
\begin{equation}
  \label{poissontwisted}
  \frac1{L^3}\sum_{\vec n}I(\vec q_n)=\frac1{L^3}\sum_{\vec n}
  \int d^3\vec q\,\,\delta^{(3)}(\vec q-\vec q_n)I(\vec q\,)=
  \sum_{\vec n}\int\frac{d^3\vec q}{(2\pi)^3}\,\,
  e^{i\vec\theta\cdot\vec n}e^{i\vec q\cdot\vec n L}I(\vec q\,)
\end{equation}
and $\Delta G_L^{\vec\theta}$ becomes
\begin{equation}
  \label{deltaGbarth}
  \Delta G_L^{\vec\theta} = -\frac1{8\pi\sqrt{s}L}\sum_{|\vec n|\ne 0}
  \frac1{|\vec n|}e^{i\vec\theta\cdot\vec n}e^{-|\vec n|\sqrt{-k^2} L}.
\end{equation}
Again, we can express $\Delta G_L^{\vec\theta}$ in terms of the L\"uscher 
zeta-function with
twisted boundary conditions, $Z_{00}^{\vec\theta}(1,\hat k^2)$, as follows,
\begin{eqnarray}
  \label{deltaG-luscher-theta}
    \Delta G_L^{\vec\theta}&=&\frac1{8\pi\sqrt{s}}\left(
    -\sqrt{-k^2} - \frac{2}{\sqrt{\pi}L} Z_{00}^{\vec\theta}(1,\hat k^2)
  \right),\\[2mm]
    Z_{00}^{\vec\theta}(1;\hat k^2)&=&\frac{1}{\sqrt{4\pi}}\,\sum_{{\vec n}\in
\mathbb{Z}^3}
  \frac{1}{\bigl(\vec n + \vec\theta/2\pi\bigr)^2-\hat k^2}\, .
\end{eqnarray}
For the particular case of $\vec\theta=(\theta,\theta,\theta)$, 
the first few terms
of the above expansion are given by
\begin{align}
  \label{deltaGbarthexp}
  \Delta G_L^{(\theta,\theta,\theta)}(M)=&-\frac1{8\pi M L}
  \Bigg[
    6\cos\,\theta\, e^{-\kappa L} + 3\sqrt{2}(1+\cos 2\theta)e^{-\sqrt{2}\kappa L}
    \nonumber\\
    &+\frac{2}{\sqrt{3}}(3\cos\theta+\cos 3\theta)e^{-\sqrt{3}\kappa L}+\cdots
  \Bigg]
\end{align}
with $\kappa = \sqrt{-k^2}$.

\subsection{Twisted boundary conditions: only one momentum shifted}
Finally, in the case of twisted boundary conditions, when only the momentum of
one of the
particles (say, particle 1) is shifted, the allowed momenta in a box are
\begin{equation}
\label{allowed-q-moving}
\vec q_1=\frac{2\pi}{L}\vec n_1+\frac{\vec\theta}{L},\qquad
\vec q_2=\frac{2\pi}{L}\vec n_2,\qquad \vec n_1,\vec n_2\in\mathbb{Z}^3\, .
\end{equation}
The particles are not in the c.m. frame any more: the c.m.
momentum is equal to $\vec P=\vec\theta/L$. 
Hence,  we have to evaluate $\Delta G_L^{\vec\theta}$
in a moving frame with momentum $\vec P$,
\begin{align}
  \label{DeltaGmf}
  \Delta G_L^{\vec\theta} =& \frac1{L^3}\sum_{\vec n}I(\vec
q_{n})-\int\frac{d^3\vec q}{(2\pi)^3}I(\vec q\,),\quad\quad
  I(\vec q\,)
=\frac1{2\omega_1\omega_2}\frac{\omega_1+\omega_2}{P_0^2-(\omega_1+\omega_2)^2},
\nonumber\\[2mm]
  \omega_1^2=&(\vec P-\vec q)^2+m_1^2,\qquad
  \omega_2^2=\vec q^2+m_2^2,\quad\quad
  \vec q_{n}=\frac{2\pi\vec n}{L},\qquad P^2=P_0^2-\vec P^2=s.
\end{align}
Again, we can approximate the integrand by~\cite{Bernard:2012bi}
\begin{equation}
  \label{Integrandaproxmf}
I(\vec q\,)= - \frac1{2P_0}\frac1{(\vec q\,')^2
  -(\vec q\,'\cdot\vec P)^2/P_0^2-\vec k^2} + \cdots,
\qquad \vec q\,' = \vec q - \mu \vec P,
\end{equation}
where $\mu = \frac1{2}\left(1-\frac{m_1^2-m_2^2}{s}\right)$, 
$\vec k$ is the momentum of the particles in the c.m. frame, 
and the dots denote exponentially
suppressed terms. 
Using the Poisson summation formula, we arrive at
\begin{align}
  \label{DeltaGformf}
  \Delta G_L^{\vec\theta}=&-\frac1{2P_0}\sum_{|\vec n|\ne 0}e^{-i\mu\vec
P\cdot\vec n L}
  \int\frac{d^3\vec q}{(2\pi)^3}\frac{e^{i\vec q\cdot\vec n L}}
  {\vec q^2-\vec k^2-\frac{(\vec q\cdot\vec P)^2}{P_0^2}}
  \\
  =&-\frac{1}{8\pi \sqrt{s} L}\sum_{|\vec n|\ne 0}\frac1{|\hat\gamma\vec n|}
  e^{-i\mu\vec\theta\cdot\vec n}e^{-|\hat\gamma\vec n|\sqrt{-k^2}L},
  \quad\hat\gamma\vec n = \gamma\vec n_{\parallel}+\vec n_\perp,
\end{align}
where $\vec n_{\parallel}$ and $\vec n_{\perp}$ are the components parallel and
perpendicular to $\vec P$ of $\vec n$, and $\gamma=P_0/\sqrt{s}$
is the relativistic gamma-factor.
Once again, we can relate $\Delta G_L^{\vec\theta}$ in this case with the L\"uscher
zeta function in the moving frame
$Z_{00}^{\vec d}(1;(q^*)^2)$~~\cite{Rummukainen}, see also
Refs.~\cite{Schierholz,Bernard:2012bi,LiLiu}:
\begin{eqnarray}
  \label{deltaG-luscher-mf}
  \Delta G_L^{\vec\theta}&=&\frac1{8\pi\sqrt{s}}\left(
    -\sqrt{-k^2} - \frac{2}{\sqrt{\pi}L\gamma} Z_{00}^{\vec d}(1;\hat k^2)
  \right),\\\nonumber\\
  Z_{00}^{\vec d}(1;\hat k^2)&=&\frac{1}{\sqrt{4\pi}}\,\sum_{{\vec r}\in P_d}
\frac{1}{{\vec r}^2-\hat k^2}\, ,
\nonumber\\[2mm]
P_d&=&\{{\vec r}=\mathbb{R}^3~|~r_\parallel=\gamma^{-1}(n_\parallel-\mu|{\vec
d}|),
~{\vec r}_\perp={\vec n}_\perp,~{\vec n}\in\mathbb{Z}^3\}\, ,
\end{eqnarray}
where ${\vec d}={\vec P}L/2\pi=\vec\theta/2\pi$.
 For the case of $\vec\theta=(\theta,\theta,\theta)$,
the first few terms in the above expansion are
\begin{align}
  \label{deltagbarmfexp}
  \Delta G_L^{(\theta,\theta,\theta)}(M) =& -\frac1{8\pi M L}
  \Bigg[
  \frac{6\sqrt{3}\cos(\mu\theta)}{\sqrt{\gamma^2+2}}e^{-\sqrt{\frac{\gamma^2+2}{3}}\,\kappa L}\nonumber\\
  &+3\sqrt{2}e^{-\sqrt{2}\kappa L}
  +\frac{3\sqrt{6}\cos(2\mu\theta)}{\sqrt{2\gamma^2+1}}
  e^{-\sqrt{\frac{2}{3}(2\gamma^2+1)}\,\kappa L}
    +\cdots
  \Bigg].
\end{align}

In the case of shallow bound states, the exponential factor $\kappa$
will
be usually quite small, so in order to reproduce accurately the full function,
one should
take several terms in the expansion for $\Delta G_L^{\vec\theta}$ above.

\end{document}